# The Wireless Train Communication Network: Roll2Rail vision

Juan Moreno García-Loygorri, Metro de Madrid S.A.; Javier Goikoetxea, CAF; Eneko Echeverría, CAF I+D;, Aitor Arriola, Iñaki Val, Ikerlan; Stephan Sand, Paul Unterhuber, DLR; Francisco del Río, Stadler Rail.

*This paper explains the main results obtained from the research carried out in the work package 2 (WP2) of the Roll2Rail (R2R) project. This project aims to develop key technologies and to remove already identified blocking points for radical innovation in the field of railway vehicles, to increase their operational reliability and to reduce life-cycle costs. This project started in May 2015 and has been funded by the Horizon 2020 program of the European Commission. The goal for WP2 is to research on both technologies and architectures to develop a new wireless Train Communication Network (TCN) within IEC61375 standard series. This TCN is today entirely wired and is used for Train Control and Monitoring System (TCMS) functions (some of them safety-related), operator-oriented services and customer-oriented services. This paradigm shift from wired to wireless means a removal of wirings implies, among other benefits, a significant reduction of life cycle costs due to the removal of cables, and the simplification of the train coupling procedure, among others.*

## Introduction

Railways are evolving very rapidly to meet the increasing demands of its users. Rolling stock is a cornerstone in this development, but there are some blocking points that need to be addressed. This is the main purpose of the EU-funded Roll2Rail (R2R) Project. Work package 2 (WP2) is focused on communication issues (which is one of the most challenging fields in railways [1]), and its final objective is to specify the requirements (and validate in a laboratory) for a wireless train communication network (TCN), at least with the same performance as the wired one. This wireless TCN would be very helpful to design, manufacture, operate and maintain trains from LCC (Life-Cycle Costs) point of view and able to avoid a significant number of failures due to broken connectors and cables. The partners in this WP were fourteen of the most relevant railway companies (see Table I), working as a single one.

The first task to meet this goal is to clarify the state-of-the-art of both radio technologies and related initiatives [2]. This review considered radio technologies and services, not only in railways, but also in the aeronautic, industry and automotive fields. Some of them have proven to be very useful in certain aspects (like GSM-R for train-to-ground communications), but they are unlikely to meet the requirements in other aspects. 3GPP LTE [3] and the IEEE 802.11 standards (commonly known as WiFi) [4] completed with deterministic communication features are two of the technologies that were identified as potential choices for the R2R project. Cognitive radio [5] is also a suitable approach that can help to achieve some of the objectives.

The structure of the paper is the following one: in section II the general requirements that a Wireless TCMS needs to meet are depicted; in section III it is provided a model for each one of the radio channels of the project; in section IV, the security and reliability, availability, maintainability, and safety (RAMS) issues are explained; in section V both the architectures and technologies are covered; in section VI, it is discussed the standardization effort carried out for train-to-ground communications; in section VII the main results are included for simulations and lab testing. Finally, the paper ends with a brief conclusion.

## General requirements

The starting point for R2R is the need to remove some blocking points in the development of the rolling stock. Regarding this WP2, the key issue is the removal of wires for train communication network (TCN), in order to have a wireless TCMS over this wireless TCN. The first task of R2R-WP2 is to specify the requirements that a wireless TCMS should fulfill to achieve the functionality, reliability, performance, safety and security levels of traditional wired TCMS solutions. The fulfillment of these requirements is an enormous challenge in every single one of the fields mentioned before. Due to the variety of train setups and configurations, the focus of the project is both on Wireless Consist[1] Networks (WLCN) and

---

[1] A consist is a single vehicle or a group of vehicles that cannot be uncoupled during normal operation.

Wireless Train[2] Backbone (WLTB), including consist-to-consist communication. Therefore, WLTBNs (WLTB Nodes) and wireless end devices (WED) will be specified as well. The requirements for the wireless TCN are the same as for the wired TCN.

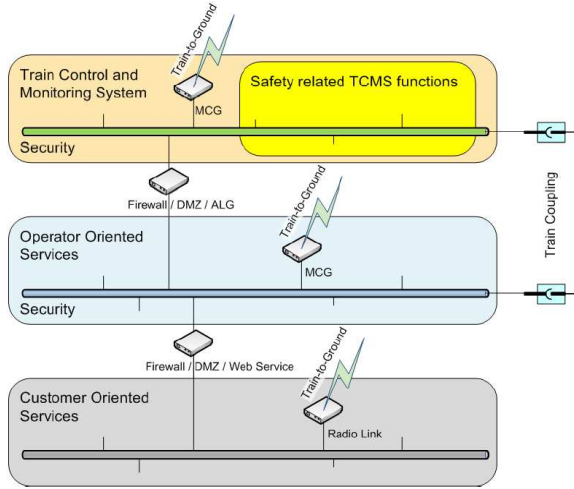

*Figure 1: Function domains*

To identify all these requirements, the CENELEC EN 15380-4 [6] functions have been checked to identify the needs of TCN regarding data exchange, and in order to map these functions into the proper domain (TCMS, operator-oriented services and customer-oriented services); see Figure 1. In other words, their required communication characteristics have been assessed. Afterwards, the wired TCN standard [7] has been reviewed to guarantee the backwards compatibility. Maintaining this compatibility is mandatory because mixed wired-wireless networks will be very common. In general, the wireless TCMS (WTCMS) is to be understood as a TCMS where the train communication network consists fully or partly of wireless networks.

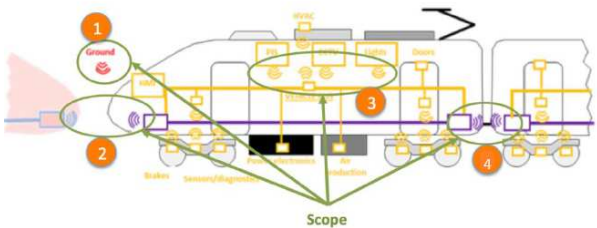

*Figure 2: Channel links within the scope of WP2 R2R Project: (1) train-to-ground, (2 consist-to-consist, (3) intra-vehicular and (4) intra-consist.*

Devices using wireless TCN can be placed in the same vehicle, consist, in different consists or even in different trains.

---

[2] A train is a composition of one or a set of consists which can be operated as an autonomous unit.

Therefore, the following communication scenarios need to be covered (see Figure 2):
- Intra-Vehicle,
- Intra-Consist,
- Consist to Consist, so called Inter-Consist,
- Train-to-Train.

If some of the coupled cars or even a consist is unable to communicate (i.e. not equipped or suffers a breakdown) the backbone communications shall not be affected.

All the train-related functions covered in [6] were classified into three different domains:
1. TCMS: includes all train control and monitoring functions (safety and non-safety) like propulsion, brakes, etc.;
2. Operator-oriented: auxiliary functions not needed for the safe movement of the train (i.e.: CCTV);
3. Customer-oriented: WiFi access, infotainment, etc.

Finally, the output from this task (the requirements' list for WLTBN, WLTB, WLCN and WED) is the main input for the other tasks.

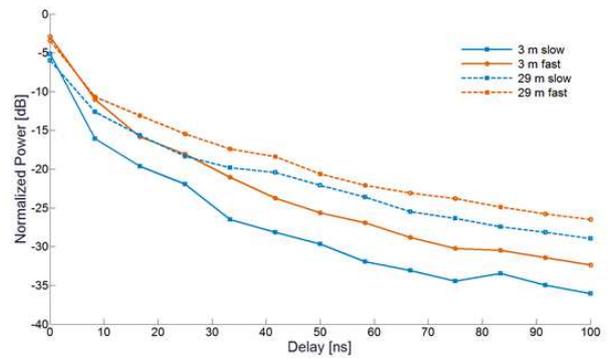

*Figure 3: Power-delay profile for inter-vehicle in HST. Two link distances at two different speeds are depicted.*

## Environmental characterization and modelling

In order to have a Wireless Train Communications Network (WTCN), it is needed to characterize the channel for radio transmission. After a thorough review of the state-of-the-art (channel models in the railway domain), we identified some gaps on this characterization of the radio environment. Unless otherwise stated, the measurements followed the 'channel sounder' methodology [8].

Due to the heterogeneity of railway scenarios, three measurement campaigns have been carried out in the field: HST (high-speed train) and metro, both of them with trains in movement, and a static 60-GHz campaign on a regional train. Several scenarios have been considered in these three campaigns (consist-to-consist, intra-consist, inter-consist, inter-vehicle; in tunnel, open air, stations, depots, etc.) to cover all the possible scenarios and obtain more general channel models. The HST measurements have been performed at 5.2 GHz, using a RUSK DLR channel sounder with a 120-MHz dynamic wideband with the train running at speeds up to 300 km/h [14].

These HST measurements have been carried out at Trenitalia facilities in Napoli, Italy. In the metro environment, a pulse-based channel sounder has been used and static as well as dynamic measurements have been performed at 2.6 GHz, using both wideband (80 MHz) and narrowband signals at 2.6 GHz, with trains running at low speeds (below 10 km/h) [15]. In this case, the narrower (80 MHz instead of 120 MHz) bandwidth means lower temporal resolution for multipath characterization (12.5 ns and 8.33 ns, respectively). These measurements have been carried out in Madrid Metro, Spain. Finally, the mmWave measurements have been performed on a static scenario at Alstom's factory in Valenciennes, France with IFSTTAR's 60-GHz channel sounder. It is noteworthy that all these scenarios are not easy to compare, because each one of them has very different circumstances (i.e. speed profiles, among others) and address different needs for the project.

In Figure 3 the HST power-delay profile (PDP) for the inter-vehicular link is shown. It can be observed that this channel has a strong line-of-sight (LOS) component, with a difference of more than 10 dB to the second tap (at 10 ns). The difference from 'slow' to 'fast' is mostly caused by different antennas in these two setups. On the other hand, this PDP has a 20-dB-power decay within 100 ns, so every tap that arrives to the receiver in this interval can be considered relevant. Figure 4 shows the PDP for the inter-consist link on a metro train in three scenarios (open field, tunnel and station). The LOS condition is maintained as well, but it can be seen that the delay spread is higher in the inter-consist link for metro than the inter-vehicular in HST. This means a lower coherence bandwidth for inter-consist links in metro trains.

All channel models obtained from the three measurement campaigns are included in the deliverable 2.2 from this WP2 [2] where all the remaining details about them can be found (i.e. antenna setups, configuration of channel sounders, etc.).

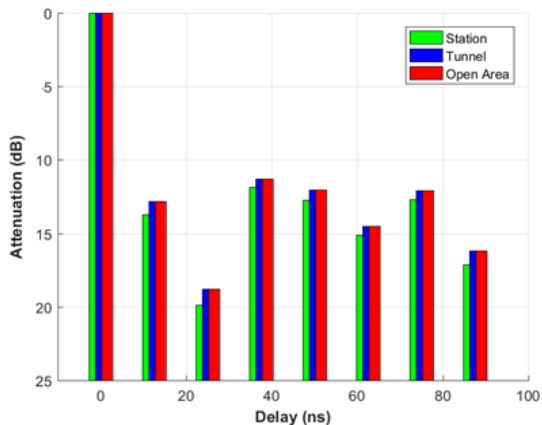

*Figure 4: Power-delay profile for inter-consist link in metro train. Three different setups have been measured: open field, tunnel and station.*

## RAMS and security analysis

The proposed wireless TCN is intended to increase the reliability, while keeping current levels of safety and security (including cyber-security) of wired systems. In first place, the possibility of having the system hacked is an important threat for wireless TCN, especially for TCMS functions. As not all kind of cyberattacks can be avoided (i.e.: jamming) the impact of cyberattacks on operational train control functions has been analyzed (what command and control subsystems would be affected and how to recognize an attack). The cybersecurity analysis has been performed following IEC 62443 standard [10].

Regarding the safety, firstly the required safety level for each one of the selected functions has been defined, so the proposed architectures and protocols could be evaluated iteratively through the standard EN50159 [11]. It has been also studied if the current SIL 2 level is achieved for TCMS functions over wireless. This calculation was based in the reliability data provided by operators and constructors, and the results agrred with the desired SIL level for the wireless TCN, which was SIL 2. In future projects we will see if these theoretical calculations agree with real-world deployments.

Finally, in an iterative process, the proposed architectures and solutions had their reliability analyzed, to detect weaknesses and suggest changes in parallel to the discussion of the proposed technologies and architectures. The first step has been to collect and classify data (both from operator's and system integrators' point of view) about reliability of existing TCMS functions in order to obtain the baseline, which will lead to the goal definition. Therefore, this RAMS & security analysis, carried out in parallel to the architecture definitions, has been essential for the success of the project.

The main results from this task are based on the (confidential) RAMS data from both operators and constructors, and we decided to keep the cyber-security risk assessment confidential, too.

## Architectures and technologies for the wireless TCMS

Once the general requirements for the wireless TCN have been clarified, it is needed to define suitable architectures and interfaces for both consist-to-consist and intra-consist communications. The proposed solutions will be inputs to the IEC TC9 WG43 in charge of the standardization of the train communication network (IEC 61375 standard [7]). The philosophy explained before has been used: three function domains (in this context it means having three networks) TCMS network, operator-oriented network and customer-oriented network. Architecture definitions will handle these three network types not only at the scope of vehicle level, but also at consist-to-consist or train-level communications.

The proposed high-level concept of the architecture is divided into two hierarchical levels: intra-consist and inter-consist, which are the train consist network and the train backbone, respectively. The element that connects both networks is the WLTBN (Wireless Train Backbone Node) and its functionality goes far beyond routing data traffic. It is noteworthy that between consists the communication must always go through the local WLTBN even if a wireless end device is able to establish a direct connection to the target consist (this communication could be safety relevant or not).

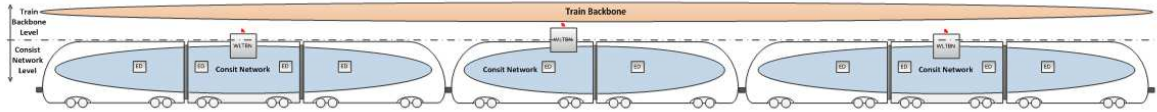

*Figure 5: General train network architecture. The Train backbone level is shown in red in the top of the figure. The consist level is shown in blue.*

Due to the fact that a train may have a dynamic network (consists can be dynamically coupled or decoupled during operation) the train integrity has to be ensured before end devices are able to communicate to each other and has to be permanently supervised. In the proposal, train integrity is achieved in two steps: in the first one the communication network is established (train discovery procedure); in the second one, the safe train-inauguration is performed. This train inauguration is safety relevant because it retrieves the orientation and sequence of vehicles that form the train. Because the number of consists in the train cannot be safely detected in the inauguration (for example, an unpowered consist at the train end), the train length has to be confirmed by another mechanism. Different strategies for the automatic train discovery procedure have been analyzed and only one of them has been finally proposed. This solution will use RFID technology and a secondary safe channel (i.e. a pneumatic pipe). 3GPP LTE has been selected as the communications technology for the train backbone (see Figure 5). We have also considered the train rescue procedure, but it is more similar to the 'dead consist' scenario, where one consist is unavailable, but the backbone connectivity needs to be maintained.

Regarding the intra-consist network, the architecture has remained more or less flexible: train lines (to achieve higher safety integrity levels when it is demanded by the function), wired consist networks and wireless networks (based on IEEE 802.11 standards) will be allowed for the implementation.

## Train-to-ground architecture and interface standardization

There is another type of communication level within the scope of the project: Train-to-Ground (T2G) communication. The T2G link is not directly covered by the R2R project, but a common and secure interface to the wayside is required to complete the train communication network concept. Therefore, the goal has been focused on contributing to close all pending clauses of the IEC 61375-2-6 draft standard[3], by providing inputs to the working group in charge of the standardization of the train communication (IEC TC9 WG43).

T2G communication system is composed by the mobile communication gateway (MCG), the ground communication gateway (GCG) –both of them with their interfaces, functions, etc. – and the wireless link. It is noteworthy that this approach does not mean the implementation of a specific technology.

To address the Open Points and group the contribution proposals, nine categories have been identified: use cases, addressing, services, protocols, security, multiple MCG handling, multiple GCG handling, multiple railway operators and the application interface. Several contribution iterations have been performed together to the IEC TC9 WG43, that are helping to complete the standard.

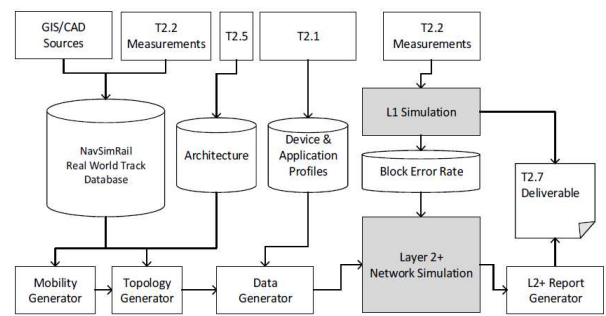

*Figure 6: general concept for the simulations.*

## Results

From all previous tasks the requirements for the Wireless TCN are obtained. In the final tasks of the project, the objective is to simulate and evaluate the selected wireless technologies and to validate these suitable radio technologies for wireless TCN.

### Simulation results

The proposed architectures and radio technologies have been simulated to check their performance in the various scenarios of the project (intra-vehicular, intra-consist, inter-consist and

---

[3] Some parts of the IEC 61375 series refer to the (wired) train backbone other parts to the (wired) consist network and one part refers to the train to ground communication, which is currently under preparation

consist-to-consist) and their suitability for the intended applications.

In a first step, a unique wireless communications technology has been preselected for simulations and validation. From our deliverable 2.3 ('State-of-the-art') [2], five candidate technologies have been selected: Ultra-wideband (UWB), the IEEE 802.11a/b/g/n/ac series of standards, intelligent transportation systems at 5.9 GHz (ITS-G5) and Long-Term Evolution (LTE). All these technologies have been evaluated against several requirements such as throughput and latency for the three different function domains, types of train communications links (i.e. intra-consist, inter-consist, T2G, etc.) and technology maturity resulting in the benefits and drawbacks summarized in Table 1 (more details on D2.7 [2]). Although all technologies have their benefits and drawbacks, LTE has been chosen for simulation and validation tasks, due to its robustness with respect to interference and the vehicular environment, applicability to T2G communications, and especially due to the near future extensions for D2D and V2X communications.

| Technology | Benefits | Drawbacks |
|---|---|---|
| UWB | More robust vs. jamming attack | Not valid for T2G, sensitive to environmental barriers |
| IEEE 802.11 | Unlicensed bands, widely used | Easily jammed, less robust to interferences, not always valid for T2G (distance dependent) |
| ITS-G5 | Already assigned license band | Also used for vehicular communications, not valid for intra-consist communications |
| 3GPP LTE | More robust to interferences, valid also for T2G, D2D and V2X communications possible in near future, evolution to 5G | License under payment unless public-safety frequency bands or a dedicated frequency band is assigned for railways |
| Millimetre Wave (60GHz) | Unlicensed band, likely evolution from LTE to 5G, high throughput | Not mature technology, sensitive to environmental barriers |

*Table 1:* Technology Selection: Benefits and drawbacks of different communications technologies for train communications

In a second step, physical-layer simulations for both LTE downlink and uplink transmissions have been conducted for different scenarios and channel models described in [9]. As an example, Fig. 7 shows the packet error rate (PER) vs. Signal-to-Noise-Ratio (SNR) for a downlink transmission, open field HSR channel and different modulations and coding schemes. The schemes R2, R3 and R9 can provide up to 7.884 Mbps, 12.586 Mbps and 55.498 Mbps net physical layer throughput at the expense of an increased SNR. For example, at a PER of 2•10-2, R3 needs 8.5 dB and R9 18 dB more SNR than R2.

In a third step, the physical layer simulation results together with the simulation scenario, the communications architecture, a device list and application profiles are then input into a layer 2+ network simulator to create realistic system level simulations. With these system level simulations, reports are generated to confirm whether the communication requirements of specific applications have been met or not. The general architecture for the simulations can be seen in Fig. 6.

*Validation*

As a final task, the proposed radio technologies and architectures have been validated in the laboratory, focusing on the wireless train backbone (WTB), which involves an inter-consist link (i.e. consist-to-consist). Several inputs have been taken from previous Roll2Rail tasks: general requirements, train channel models (i.e. tunnel, station and open field channels for both metro and high-speed trains), and architectures. A comparison with simulation results in terms of Packet Error Rate (PER) has been done too.

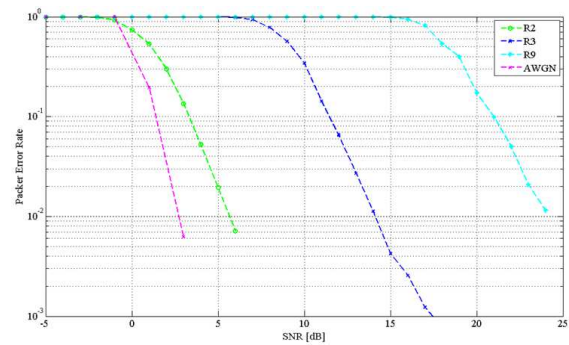

*Figure 7:* Packet Error Rate vs. SNR [dB]: LTE downlink transmission for open field high-speed rail channel and different modulation and coding schemes. R2, R3 and R9 stand for 7.884 Mbps, 12.586 Mbps and 55.498 Mbps net physical layer throughput. The AWGN curve is plotted as a reference. R3 and R9 are reference channels for the downlink and R2 is for the uplink, as it is in the 3GPP LTE technical specifications 36.101 and 36.104, respectively.

The central elements for the validation have been a wireless channel emulator [12] and three LTE modules. The LTE modules have been two TWS8100 UEs and one LTE-eNB composed by an Evolved Packet Core (EPC) and a TWS 8210 Remote Radio Head (RRH). These LTE devices have been operating on LTE Band 28 (703-803 MHz), in Frequency Division Duplex (FFD) mode, and configured in single-input single-output (SISO) mode. They have been connected via Ethernet ports to the railway communication equipment. A diagram and a picture of the validation setup are shown in Fig. 8 and 9, respectively.

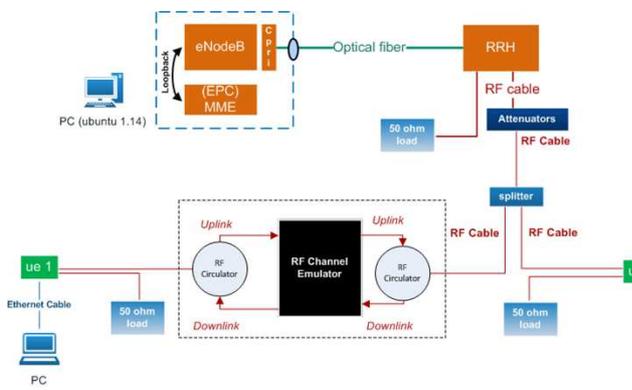

*Figure 8: Connection of LTE devices and channel emulator in SISO mode*

Using this validation scenario, specific functionalities of the train backbone have been validated, including train inauguration and data exchange. The inauguration process is divided into two parts: first, the wireless train-backbone nodes (WLTBNs) perform the physical and logical discovery of the train topology and generate the 'train network directory', as defined in IEC 61375-2-5 standard. Once this phase ends successfully, it is followed by the TCN inauguration as defined in IEC 61375-2-3, which establishes the active cabin. When this operation is completed, inauguration can be inhibited in order to avoid unwanted topology changes due to connection failures. Finally, once this operational inauguration ends, the WLTBNs generate the 'operational train directory'. The previous inauguration process has been validated using Router PCs and EBN-600 Ethernet Train Backbone Nodes (ETBN, developed for this project) connected to the wireless scenario. This setup emulated a train formed by three consists each one composed by four cars. Regarding the validation of the data-exchange functionality, railway-specific end devices have been connected, such as input-output (IO) modules, human-machine interface (HMI) devices and closed circuit television (CCTV) cameras.

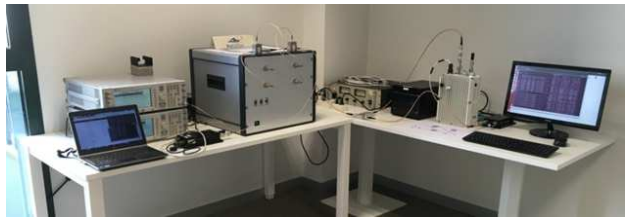

*Figure 9: Validation Scenario*

| Partner | Type of company | Country |
|---|---|---|
| ALSTOM | Train manufacturer | France |
| BOMBARDIER | Train manufacturer | Germany |
| CAF | Train manufacturer | Spain |
| DEUTSCHE BAHN | Railway operator | Germany |
| DLR | Research instit. | Germany |
| HITACHI RAIL ITALY | Train manufacturer | Italy |
| IFSTTAR | Research instit. | France |
| IK4- IKERLAN | Research instit. | Spain |
| METRO DE MADRID | Railway operator | Spain |
| SIEMENS | Train manufacturer | Germany |
| SNCF | Railway operator | France |
| STADLER | Train manufacturer | Spain |
| THALES | Subsystem supplier | France |
| TRENITALIA | Railway operator | Italy |
| UNICONTROLS | Subsystem supplier | Czech Rep. |
| UNIV. OF SALZBURG | Research instit. | Austria |

*Table 2: R2R WP2 partners. CAF is both the R2R Technical Leader and the WP2 leader.*

## Conclusions

In this paper a very brief summary of the main results achieved by the WP2 of the EU-funded Roll2Rail project has been presented. This WP2 is focused on the development of a Wireless TCN to carry all the train-related data –but the signaling one-: TCMS, operator-oriented and customer-oriented services.

A brief specification of the requirements has been provided (a more detailed can be found in the deliverable 2.1 of the project [2]); three measurements campaigns to obtain the needed channel models to characterize the required radio transmissions have been performed; a RAMS and security analysis; the architectural design for both the intra-consist and train backbone networks; simulations and validation in the laboratory.

The results and findings of the project have been transferred to the larger Shift2Rail initiative [13], whose projects focused on TCMS will end in 2022 with the implementation of full scale demonstrators including the wireless TCMS here described.

**Acknowledgements**

The authors are thankful for the support of the European Commission through the Roll2Rail Project, one of the lighthouse projects of the Shift2Rail initiative under the Horizon 2020 programme. The Roll2Rail Project has received funding from the European Union's Horizon 2020 research and innovation programme under the Grant Agreement no. 636032.

**Juan Moreno García-Loygorri** (MSc 2007, PhD 2015) works as a rolling stock engineer in the Engineering and Research Department of Metro de Madrid, and also as a part-time professor in the Universidad Politécnica de Madrid. He has been working in railways since 2007. He has participated in many railway-related research projects and has authored more than 30 papers on railway communications. His research interests are channel measurement & modelling, railway communications systems and software-defined radio.

**Javier Goikoetxea** has been working as Project Coordinator at the Technology Division of CAF since 1997. Javier has been deeply involved in the development of CAF's TCMS product. Today he is the Technical Leader of Roll2Rail and also leads the TCMS activities within the Shift2Rail initiative. He is the chairman of UNIFE's Topical Group TCMS and a Spanish national expert at the IEC TC9 WG43.

**Eneko Echeverria** graduated in Telecommunications Engineering in 2006. Soon after, he joined the research division at CAF for 1 year. After it he worked for CAF Power and Automation in the electronic department during three years. Finally, in 2010 he went back to the research division at CAF, where he started his role as validation and verification manager for the new generation of safety-related electronic platforms of CAF Group, and nowadays he shares that responsibility with the role of WP2 leader of Roll2Rail.

**Aitor Arriola** received the M.S. degree in telecommunications engineering from the University of the Basque Country (UPV/EHU), Bilbao, Spain, in 1999, and the Ph.D. degree from the University of Navarra, San Sebastian, Spain, in 2011. Since 1999, he has been with the Communications Systems Group of IK4-IKERLAN Technological Research Center, Arrasate-Mondragón, Spain. From 2006 to 2008, he was with the RF Component Design and Modeling (RFCDM) Group of Imec, Leuven, Belgium, as visiting researcher. He has participated in several national and international research projects involving the design and measurement of wireless communication devices and antennas.

**Iñaki Val** received B.S. and M.S. degrees from Mondragon University (Spain) in 1998 and 2001, respectively, and a Ph.D. from the Universidad Politécnica de Madrid in 2011. Since 2001, he has been with the Communications Systems Group of IK4-IKERLAN, and in the past has been with Fraunhofer IIS of Erlangen (Germany) as invited researcher (2005-2006). Currently, he is the Team Leader of Communication Systems group. His research activities include the design and implementation of digital wireless communications systems, cognitive radio, wireless channel characterization and digital signal processing. He is currently focused on industrial communications applications.

**Stephan Sand** received the PhD degree from the Swiss Federal Institute of Technology (ETH) Zurich, Switzerland in 2010. In 2002, he joined the Institute of Communications and Navigation of the German Aerospace Center (DLR) in Oberpfaffenhofen, Germany. Currently, he is leading the Vehicular Applications Group and researching novel systems that combine robust navigation and wireless communications technologies. These systems will protect vulnerable road users, e.g. bicyclist and pedestrians, and increase traffic efficiency and safety in railways. Stephan has authored and co-authored more than 100 technical and scientific publications in conferences and journals and obtained several patents on his inventions.

**Paul Unterhuber** received the Dipl-Ing. degree in electrical engineering and in electrical engineering and business from Graz University of Technology in 2013 and 2014. In 2015, he joined the Institute of Communications and Navigation of DLR in Oberpfaffenhofen, Germany. He has been researching the propagation of wireless communication signals between trains and in particular measuring and modelling railway communication channels. As a member of the Vehicular Applications Group he led, managed, and executed the world wide first train-to-train measurements for high-speed trains within the EU Horizon 2020 project Roll2Rail.

**Francisco del Río** is working as Electronics and Software development Manager at STADLER Valencia, where has been working since 2007. Francisco is leading the development of TCMS, Multimedia Systems and Telemetry solutions for light rail vehicles and locomotives at STADLER's Valencia site, He is a Telecommunication Engineer.